\documentclass[aps,prd,twocolumn,longbibliography,nofootinbib]{revtex4-2}

\usepackage{amsmath,amssymb,amsfonts}
\usepackage{graphicx}
\usepackage{slashed}
\usepackage{dsfont}
\usepackage{xcolor}
\usepackage{hyperref}

\usepackage{array}
\usepackage{scalerel}
\usepackage{extarrows}
\usepackage{arydshln}

\newcommand{\be}{\begin{equation}}
\newcommand{\ee}{\end{equation}}
\newcommand{\bea}{\begin{eqnarray}}
\newcommand{\eea}{\end{eqnarray}}
\newcommand{\nn}{\nonumber}

\begin{document}

\title{Kerr–Schild Double Copy of the Randall–Sundrum Black String}

\author{Jesús A. Rodríguez}
\email{jesurodriguez@uade.edu.ar}
\affiliation{Universidad Argentina de la Empresa (UADE), Facultad de Ingeniería y Ciencias Exactas, Departamento de Ciencias Básicas, Lima 717, Buenos Aires, Argentina}

\begin{abstract}

We construct the Kerr–Schild classical double copy of the black string in the Randall–Sundrum II model, deriving the single and zeroth copies, and verifying the associated field equations. The single copy gauge field is independent of the holographic coordinate and satisfies a sourceless Maxwell equation on the curved background, in direct analogy with the Coulomb field of the Schwarzschild double copy. The zeroth copy scalar obeys a modified Klein--Gordon equation with a first-order derivative term along the extra dimension; a field redefinition yields a standard Klein--Gordon equation with effective mass $m^2 = 12/l^2$, induced by the warp factor. We further show that an alternative Kerr--Schild splitting, gravitationally equivalent to the canonical one, produces a physically inequivalent double copy: the gauge field is supported by a conserved but delocalized bulk current, and the zeroth copy satisfies a massless equation that carries no imprint of the warped extra dimension.
   
\end{abstract}
                   

\maketitle

\section{Introduction}
\label{sec:intro}

The relationship between gauge theories and gravity has long been a central theme in theoretical physics, appearing in contexts ranging from the AdS/CFT correspondence to the more recent double copy paradigm. The latter emerged from the Bern–Carrasco–Johansson (BCJ) color-kinematics duality in scattering amplitudes \cite{Bern:2008qj,Bern:2010ue}, which establishes that gravitational amplitudes can be systematically constructed from gauge-theory building blocks, schematically expressed as
\be
\text{Gravity} = \text{(Gauge Theory)} \times \text{(Gauge Theory)}\, \nn .
\ee
Although originally formulated for perturbative scattering amplitudes, this correspondence has been successfully extended to exact classical solutions \cite{Monteiro:2014cda}, where spacetimes admitting a Kerr–Schild structure \cite{Kerr:1965wfc} possess a natural gauge counterpart.

In the Kerr–Schild double copy framework, the gravitational metric is written as
\be
\label{eq:KS-ansatz}
g_{MN} = \bar{g}_{MN} + \phi\,k_{M}k_{N}\, ,
\ee
where $\bar{g}_{MN}$ is a background metric (typically flat or maximally symmetric), $\phi$ is a scalar function, and $k^{M}$ is a null vector field that is geodesic with respect to both the background and the full metric. A key feature of this ansatz is the linearization of the Einstein equations, enabling a direct identification of gauge-theory solutions. The single copy corresponds to the gauge field $A_M = \phi\, k_M$, satisfying Maxwell equations, while the zeroth copy is the scalar $\Phi=\phi$, obeying the corresponding scalar equation. The Kerr--Schild double copy has been extensively explored for a wide class of stationary, cosmological, and extended solutions \cite{Luna:2015paa, Bahjat-Abbas:2017htu,Carrillo-Gonzalez:2017iyj,Lee:2018gxc,Cho:2019ype,Lescano:2020nve,Berman:2020xvs,Lescano:2021ooe,Alkac:2021bav,Angus:2021zhy,Lescano:2022nhp,Berman:2022bgy,Rodriguez:2025hiu,Alencar:2026zdz,Morieri:2026gdo}.

An important subtlety of the Kerr--Schild ansatz is that the decomposition~\eqref{eq:KS-ansatz} is not unique. The rescaling
\be
\label{eq:ambiguity}
k_{M} \;\rightarrow\; a(x)\,k_{M}\,, \qquad
\phi \;\rightarrow\; \frac{\phi}{a(x)^2}\,,
\ee
leaves the metric invariant while producing different gauge-theory fields. As emphasized in \cite{Carrillo-Gonzalez:2017iyj}, this ambiguity must be resolved by imposing physical criteria on the single copy, such as the absence of delocalized sources. Different splittings that are gravitationally equivalent can thus yield physically inequivalent gauge-theory descriptions, making the resolution of this ambiguity a non-trivial and physically meaningful step in the double copy construction.

A complementary line of research in gravitational physics concerns the possibility of extra spatial dimensions. This idea received renewed attention at the turn of the century with the Randall–Sundrum models \cite{Randall:1999ee,Randall:1999vf}, in which our four-dimensional universe is realized as a hypersurface (brane) embedded in a five-dimensional AdS bulk. Of particular relevance to the present work is the RSII model \cite{Randall:1999vf}, where gravity is effectively localized on the brane even though the extra dimension is infinite and non-compact, provided the bulk geometry is appropriately warped \cite{Garriga:1999yh}. In this setup, the metric takes the form
\begin{equation}
\label{eq:RSII-metric}
ds^2 = e^{-2|y|/l}\eta_{\mu\nu}dx^\mu dx^\nu + dy^2\,,
\end{equation}
where $l$ is the AdS$_5$ curvature radius and $y \in (-\infty, +\infty)$ is the extra-dimensional coordinate with a $\mathbb{Z}_2$ orbifold symmetry. The exponential warp factor renders the four-dimensional graviton zero mode normalizable, reproducing Newtonian gravity at large distances together with calculable short-distance corrections.

Black hole solutions incorporating the effects of the warped extra dimension have also been extensively studied \cite{Chamblin:1999by,Dadhich:2000am,Tanaka:2002rb,Emparan:2002px,Aliev:2005bi,Amarilla:2011fx, Neves:2012it}, exhibiting genuine extra-dimensional features absent in standard four-dimensional general relativity. In particular, Ref.~\cite{Chamblin:1999by} showed that the Schwarzschild solution can be extended into the bulk as a \emph{black string}, obtained by uniformly extending the four-dimensional geometry along the extra dimension. 

Despite these developments, the classical double copy in the presence of warped extra dimensions remains, to the best of our knowledge, unexplored. The Randall–Sundrum framework provides a natural setting to address this question, combining AdS geometry, gravity localization, and exact black hole solutions that admit a Kerr–Schild representation. Understanding how characteristic brane-world features are encoded in gauge-theory data may offer new insights into both the double copy correspondence and extra-dimensional gravity.

In this work, we initiate the study of the classical double copy for black holes in Randall–Sundrum braneworlds by focusing on the RSII black string \cite{Chamblin:1999by}. We derive the corresponding single and zeroth copies for the canonical Kerr--Schild splitting of~\cite{Carrillo-Gonzalez:2017iyj} and verify the associated field equations, showing that the RSII warp factor leaves a direct imprint on the gauge-theory description. A central result of this work is that the ambiguity inherent in the Kerr--Schild decomposition is physically consequential in the warped setting: we construct an alternative splitting that is gravitationally equivalent to the canonical one but yields a physically inequivalent double copy, with a gauge field supported by a delocalized bulk current and a zeroth copy that carries no imprint of the warped extra dimension.

\section{Setup and theoretical framework}
\label{sec:setup}

In this section we collect the technical ingredients underlying our analysis. We first review the Kerr--Schild double copy in curved backgrounds, focusing on the linearized field equations in the presence of a cosmological constant, and then present the RSII model and the black string solution that will serve as the gravitational starting point. We denote bulk indices by $M, N, \ldots = 0, 1, \ldots, D{-}1$, while four-dimensional brane indices are denoted $\mu, \nu, \ldots = 0, 1, 2, 3$. For the RSII model $D=5$.

\subsection{Classical double copy in curved backgrounds}
\label{subsec:double-copy}

The Kerr--Schild ansatz and the rescaling ambiguity were introduced in equations \eqref{eq:KS-ansatz} and \eqref{eq:ambiguity}. We collect here the additional properties needed in subsequent sections.

The nullity of $k^M$ ensures that the inverse metric $g^{MN} = \bar{g}^{MN} - \phi\, k^{M}k^{N}$, remains linear in $\phi$, and that the determinant satisfies $\det(g) = \det(\bar{g})$. As a consequence, the Einstein equations linearize exactly in the perturbation $h_{MN} = \phi\, k_M k_N$. For a background $\bar{g}_{MN}$ solving Einstein's equations with cosmological constant $\Lambda_D$, the linearized vacuum equations for the Kerr--Schild perturbation take the form~\cite{Bahjat-Abbas:2017htu,Carrillo-Gonzalez:2017iyj}
\be
\label{eq:app-linear-eom-lambda}
\bar{\nabla}_{P}\!\left[2\bar{\nabla}_{(M}h_{N)}{}^{P}
- \bar{\nabla}^{P}h_{MN}\right]
=
\frac{4\Lambda_{D}}{D-2}h_{MN}\,.
\ee

The gauge-theory copies are extracted by contracting this tensorial equation with Killing vectors of both the exact geometry and the background. Given a Killing vector $\xi^M$ of both $g_{MN}$ and $\bar{g}_{MN}$:

\medskip
\noindent\textbf{Single copy:} A single contraction of \eqref{eq:app-linear-eom-lambda} with $\xi^M$, combined with the Killing equation and the null geodesic properties of $k_M$, yields a Maxwell-type equation for the gauge field
\be
\label{eq:app-single}
A_M = \phi\, k_M\,.
\ee
The explicit form of this equation depends on the background geometry and the choice of Killing vector.

\medskip
\noindent\textbf{Zeroth copy:} A further contraction with the same Killing vector produces a scalar equation for
\be
\label{eq:app-zeroth}
\Phi = \phi\,.
\ee

In flat backgrounds ($\Lambda_D = 0$, $\bar{g}_{MN} = \eta_{MN}$), these reduce to the sourceless Maxwell and Klein--Gordon equations, respectively~\cite{Monteiro:2014cda}. In curved backgrounds, the cosmological constant and the non-trivial connection generally modify both equations, as we shall see in detail for the RSII geometry. Moreover, the rescaling freedom \eqref{eq:ambiguity} leaves the gravitational solution invariant but modifies the gauge-theory fields. We will exploit this freedom to construct both a canonical and an alternative realization of the double copy, and show that they lead to physically distinct gauge-theory descriptions.

\subsection{The RSII black string} 
\label{subsec:black-string}

The RSII model describes a single positive-tension 3-brane embedded in an AdS$_5$ bulk, with the metric given by \eqref{eq:RSII-metric}. Here $\eta_{\mu\nu}$ is the four-dimensional Minkowski metric, $y \in (-\infty,+\infty)$ is the extra-dimensional coordinate, and $l$ is the AdS$_5$ curvature radius, related to the bulk cosmological constant by $\Lambda_{5} = -6/l^2$. The brane is located at $y=0$ and a $\mathbb{Z}_2$ reflection symmetry $y \leftrightarrow -y$ is imposed, so that the geometry on both sides of the brane is identical. The brane tension $\sigma$ is related to $l$ and the five-dimensional Newton constant $G_5$ through the Israel junction conditions \cite{Israel:1966rt,Chamblin:1999ya} by $\sigma = 3/(4\pi G_5 l)$. A key property of this background is the exponential warp factor $e^{-2|y|/l}$, which localizes the four-dimensional graviton zero mode near the brane \cite{Randall:1999vf}.

It will be convenient to introduce the conformal coordinate $z = l\,e^{|y|/l}$, in terms of which the metric \eqref{eq:RSII-metric} becomes
\be
\label{eq:AdS5-conformal}
ds^{2} = \frac{l^2}{z^2}\left(
\eta_{\mu\nu}\,dx^\mu dx^\nu + dz^2\right)\,.
\ee
Since $|y| \geq 0$, the conformal coordinate satisfies $z \geq l$, with the brane at $z = l$ and the Poincar\'e horizon at $z \to \infty$. The $\mathbb{Z}_2$ orbifold identifies the two sides of the brane, and we work on the fundamental domain $z \in [l, \infty)$. When needed for the Kerr--Schild construction, we extend the background metric \eqref{eq:AdS5-conformal} to the full Poincar\'e patch $z \in (0, \infty)$, which includes the AdS$_5$ boundary at $z \to 0$.

A central property underlying the construction of black hole solutions in this background is that the metric~\eqref{eq:AdS5-conformal} is a conformal rescaling of flat space, so the five-dimensional Einstein equations with $\Lambda_5 = -6/l^2$ are automatically satisfied whenever the seed metric replacing $\eta_{\mu\nu}$ is Ricci flat~\cite{Brecher:1999xf}. Replacing $\eta_{\mu\nu}$ with the four-dimensional Schwarzschild metric therefore yields an exact solution known as the black string~\cite{Chamblin:1999by}
\be
\label{eq:black-string-z}
ds^2 = \frac{l^2}{z^2}\left(-f(r)\,dt^2 
+ f(r)^{-1}\,dr^2 + r^2\,d\Omega_2^2 + dz^2\right)\,,
\ee
where $f(r) = 1 - \frac{2M}{r}$, and $d\Omega_2^2 = d\theta^2 + \sin^2\theta\,d\varphi^2$ is the metric on the unit two-sphere. The induced metric on the brane at $z = l$ is the four-dimensional Schwarzschild solution, consistent with the spherical symmetry of the seed. The solution describes a horizon at $r = 2M$ that extends uniformly along the extra dimension, forming a string-like object in the five-dimensional bulk.

Although this is an Einstein metric, so that the Ricci scalar and the square of the Ricci tensor remain finite everywhere, the Kretschmann scalar 
\be
\label{eq:kretschmann}
R_{MNPQ}R^{MNPQ} = \frac{1}{l^4}\left(40 
+ \frac{48M^2 z^4}{r^6}\right)
\ee
diverges both at $r = 0$ (the black string singularity) and as $z \to \infty$ (the Poincar\'e horizon), signalling a curvature singularity at the AdS$_5$ horizon. Moreover, the black string is subject to the Gregory--Laflamme instability \cite{Gregory:1993vy} for sufficiently small masses. The endpoint of this instability is expected to be a stable ``black cigar'' that coincides with the black string in the vicinity of the brane but whose horizon closes off before reaching $z \to \infty$ \cite{Chamblin:1999by}. 

Despite these limitations, the black string \eqref{eq:black-string-z} provides an exact and analytically tractable solution that is well-suited for studying the Kerr--Schild double copy in braneworld scenarios. In particular, its warped structure will play a central role in what follows.

\section{Classical double copy of the RSII black string} 
\label{sec:double-copy}

In this section we construct the classical double copy of the RSII black string \eqref{eq:black-string-z}. We recast the metric in Kerr--Schild form over the AdS$_5$ background, derive the associated single and zeroth copies following the prescription of~\cite{Carrillo-Gonzalez:2017iyj}, and verify the corresponding field equations on the curved background. We then consider an alternative Kerr--Schild splitting and show that it leads to a physically inequivalent double copy.

\subsection{Kerr--Schild decomposition}
\label{subsec:KS-decomp-CPT}

To bring the RSII black string~\eqref{eq:black-string-z} into Kerr--Schild form over the AdS$_5$ background~\eqref{eq:AdS5-conformal}, we introduce ingoing Eddington--Finkelstein coordinates
\be
\label{eq:EF-coord}
dv = dt + \frac{dr}{f(r)}\, ,
\ee
in terms of which the metric becomes
\be
\label{eq:black-string-EF}
ds^2 = \frac{l^2}{z^2}\left(-dv^2 + 2\,dv\,dr + r^2\,d\Omega_2^2 + dz^2 + \frac{2M}{r}\,dv^2\right)\, .
\ee
The first four terms are the AdS$_5$ background in Eddington--Finkelstein form, while the last is a rank-one perturbation. The metric therefore admits the Kerr--Schild decomposition \eqref{eq:KS-ansatz} with
\be
\label{eq:KS-data}
\phi = \frac{2M}{r}\,\frac{z^2}{l^2}\,, \qquad
k_M\,dx^M = \frac{l^2}{z^2}\,dv\, .
\ee

We now verify the required properties of $k_M$. The vector has a single non-vanishing component $k_{v}$. Writing the background as $\bar{g}_{MN} = (l^2/z^2)\,\hat{g}_{MN}$, where $\hat{g}_{MN}$ denotes the flat five-dimensional metric in Eddington--Finkelstein coordinates, the contravariant components are
\be
\label{eq:k-contra}
k^M = \bar{g}^{MN}k_N = \hat{g}^{Mv} = \delta^M_r\,,
\ee
where we used $\hat{g}^{vv} = 0$ and $\hat{g}^{rv} = 1$. Nullity
follows immediately
\be
\bar{g}_{MN}\,k^M k^N = \bar{g}_{rr} = \frac{l^2}{z^2}\,\hat{g}_{rr} = 0\,.
\ee

For the geodesic property, we compute
\be
k^N\bar{\nabla}_N k_M = \partial_{r}\left(\frac{l^2}{z^2}\delta_{M}^{v}\right) - \bar{\Gamma}_{rM}^{v}\frac{l^2}{z^2}\, .
\ee
The first term vanishes as the only non-zero component of $k_{M}$ is independent of $r$. For the second term, it can be shown that $\bar{\Gamma}^{v}_{rM} = 0$ for all $M$. Therefore $k^M$ is an affinely parametrized null geodesic of the AdS$_5$ background.

\subsection{Single copy}
\label{subsec:single-copy-CPT}

Having obtained the Kerr--Schild decomposition of the RSII black string, the single copy gauge field is defined through \eqref{eq:app-single} and the identifications \eqref{eq:KS-data} as 
\be
\label{eq:A-single}
A_{M} = \phi k_{M} = \frac{2M}{r}\delta_{M}^{v}\, ,
\ee
so that the only non-vanishing component is
\be
\label{eq:A-v}
A_v = \frac{2M}{r}\, .
\ee
The field strength tensor, $F_{MN} = \partial_M A_N - \partial_N A_M$, has a single independent non-vanishing component,
\be
\label{eq:F-vr}
F_{vr} = -\partial_r A_v = \frac{2M}{r^2}\,,
\ee
with all other components vanishing. In particular, the gauge field and its field strength are independent of the bulk coordinate $z$, reflecting the fact that the warp factor has been fully absorbed into the Kerr--Schild decomposition.

The single copy equation of motion is obtained by contracting the linearized Einstein equation \eqref{eq:app-linear-eom-lambda} with a Killing vector $\xi^M$ of the background. We choose $\xi = \partial_v$, which satisfies $\xi^M k_M = l^2/z^2$, naturally aligned with the direction along which $k_M$ is supported. Specializing to RSII ($D = 5$, $\Lambda_5 = -6/l^2$), the contraction gives
\be
\label{eq:contraction-step1}
\xi^N\bar{\nabla}_P\!\left[2\bar{\nabla}_{(M}h_{N)}{}^{P} - \bar{\nabla}^{P}h_{MN}\right] = -\frac{8}{z^2}\,\phi\,k_M\, .
\ee
Defining $T_{MNP} = 2\bar{\nabla}_{(M}h_{N)P} - \bar{\nabla}_{P}h_{MN}$, this can be written as
\be
\label{eq:contraction-split}
\bar{g}^{PQ}\bar{\nabla}_P\!\left(\xi^N T_{MNQ}\right) - \bar{g}^{PQ}\!\left(\bar{\nabla}_P\xi^N\right)T_{MNQ} = -\frac{8}{z^2}\,A_M\, ,
\ee
where we identified $\phi\,k_M = A_M$ and moved $\xi^N$ inside the covariant derivative, thus introducing a connection term.

Using adapted coordinates in which $\xi^M = \delta^M_v$ is a coordinate basis vector and evaluating the Christoffel symbols for the AdS$_5$ background, eq. \eqref{eq:contraction-split} reduces to
\be
\bar{g}^{PQ}\bar{\nabla}_P\!\left[
\frac{l^2}{z^2}\,F_{MQ}
+ \frac{2l^2}{z^3}\,\delta_Q^z\,A_M\right]
= -\frac{10}{z^2}\,A_M\,.
\ee
The second term inside the brackets can be evaluated directly
\be
\bar{g}^{PQ}\bar{\nabla}_P\!\left[
\frac{2l^2}{z^3}\,\delta_Q^z\,A_M\right]
= -\frac{10}{z^2}\,A_M\,,
\ee
which exactly cancels the right-hand side, leaving
\be
\label{eq:Maxwell-warped}
\bar{\nabla}_N\!\left(\frac{l^2}{z^2}\,F_M{}^N\right) = 0\,.
\ee

Expanding the covariant divergence using
$\bar{\nabla}_N(l^2/z^2) = -(2l^2/z^3)\,\delta_N^z$, this can be
equivalently written as
\be
\label{eq:Maxwell-expanded}
\bar{\nabla}_N F_M{}^N - \frac{2}{z}\,F_M{}^z = 0\,,
\ee
making explicit the effect of the AdS$_5$ warp factor on the gauge
dynamics through the coupling to $F_M{}^z$. Since $F_{MN}$ has support only along the $(v,r)$ directions, raising indices with the background metric does not generate a $z$-component 
\be
F_M{}^z = \bar{g}^{zP}F_{MP} = \frac{z^2}{l^2}\,F_{Mz} = 0\,,
\ee
and thus the second term in \eqref{eq:Maxwell-expanded} vanishes identically. The equation of motion reduces then to
\be
\label{eq:Maxwell-final}
\bar{\nabla}_{N}F_{M}{}^{N}=0\,,
\ee
away from the singularity at $r=0$.

This equation must be understood in a distributional sense. The gauge field \eqref{eq:A-v} is singular at $r=0$, and therefore the divergence $\bar{\nabla}_{N}F_{M}{}^{N}$ vanishes everywhere except at the origin, where it produces a localized delta-function source. In this sense, the solution describes a point-like field generated by a localized charge at $r=0$, in direct analogy with the Coulomb field arising from the double copy of the Schwarzschild geometry \cite{Monteiro:2014cda}.

It is worth emphasizing that the $z$-independence of $A_v$ follows directly from the Kerr--Schild decomposition and should not be interpreted as a dynamical localization of the gauge field on the brane: gauge fields cannot be gravitationally trapped in the RSII scenario~\cite{Davoudiasl:1999tf,Pomarol:1999ad}, and their confinement requires additional non-gravitational mechanisms \cite{Ghoroku:2001zu,Alencar:2026afq}. From the braneworld perspective, the restriction to the brane at $z = l$ reproduces the standard four-dimensional single copy, consistent with the role of the brane as the hypersurface where four-dimensional physics is recovered~\cite{Garriga:1999yh}.

\subsection{Zeroth copy}
\label{subsec:zeroth-copy-CPT}

The final step in the classical double copy chain is to construct the zeroth copy, corresponding to a scalar field theory. This scalar is identified with the Kerr--Schild scalar function \eqref{eq:KS-data}
\be
\Phi = \phi = \frac{2M}{r}\,\frac{z^2}{l^2}\, .
\ee

The equation of motion is obtained by contracting the linearized Einstein equation \eqref{eq:app-linear-eom-lambda} twice with the same Killing vector $\xi = \partial_v$ used in the single copy construction. Using the tensor $T_{MNP}$ defined in Section \ref{subsec:single-copy-CPT}, the double contraction gives
\be
\bar{g}^{PQ}\bar{\nabla}_P\!\left(\xi^M\xi^N T_{MNQ}\right) - \bar{g}^{PQ}\bar{\nabla}_P\!\left(\xi^M\xi^N\right)T_{MNQ} = -\frac{8l^2}{z^4}\,\Phi\,.
\ee
Evaluating both terms on the left-hand side using adapted coordinates and the Christoffel symbols of the AdS$_5$ background, as in the single copy derivation, one arrives at
\be
\label{eq:scalar-warped}
\bar{\Box}\,\Phi - \frac{8z}{l^2}\,\partial_z\Phi + \frac{20}{l^2}\,\Phi = 0\, ,
\ee
away from the singularity at $r = 0$, where a localized delta-function source appears, as in the single copy case.

Equation \eqref{eq:scalar-warped} differs from the standard Klein--Gordon equation by the presence of a first-order derivative term along the holographic direction~$z$. This term has no counterpart in the flat-space or pure (A)dS zeroth copy equations
of \cite{Monteiro:2014cda,Carrillo-Gonzalez:2017iyj} and reflects the explicit $z$-dependence of the zeroth copy scalar $\Phi \sim z^2$: unlike the single copy gauge field, the coupling to the conformal factor of the RSII spacetime is no longer trivial.

It is natural to ask whether a field redefinition can bring equation \eqref{eq:scalar-warped} into standard Klein--Gordon form. Defining
\be
\Theta \equiv \frac{l^4}{z^4}\,\Phi\, ,
\ee
a direct computation shows that the first-derivative coupling is removed and \eqref{eq:scalar-warped} transforms into
\be
\label{eq:scalar-KG}
\left(\bar{\Box} - \frac{12}{l^2}\right)\Theta = 0\,.
\ee
The redefined field $\Theta$ therefore satisfies a Klein--Gordon equation with effective mass $m^2 = 12/l^2$, induced by the AdS$_5$ warp factor.

The two scalar fields have complementary bulk profiles and physical roles. The canonical zeroth copy $\Phi \sim z^2$ grows towards the Poincar\'e horizon, signalling that it is not localized near the brane and should be understood only as the geometrically natural variable directly encoding the Kerr--Schild structure. In contrast, $\Theta \sim z^{-2}$ decays along the holographic direction and is normalizable with respect to the natural bulk norm\footnote{For a scalar field in AdS$_5$ with metric \eqref{eq:AdS5-conformal}, the natural norm is $\int_l^\infty dz\,\sqrt{-\bar{g}}\,|\Theta|^2 \sim
\int_l^\infty dz\,(l/z)^5\,|\Theta|^2$. For $\Theta \sim z^{-2}$ this integral converges, while for $\Phi \sim z^2$ it diverges.}, in line with the expected behavior of normalizable bulk modes in the RSII setup \cite{Randall:1999vf,Garriga:1999yh}. The fact that $\Theta$ satisfies a Klein--Gordon equation with a constant mass further supports its interpretation as the physically propagating scalar degree of freedom in the bulk.

\subsection{Alternative Kerr-Schild splitting}
\label{subsec:warped}

Motivated by the ambiguity in the Kerr--Schild decomposition discussed in the introduction, we consider an alternative splitting of the RSII black string,
\be
\label{eq:warped-KS-data}
\phi = \frac{2M}{r}\, , \qquad k_M\,dx^M = \frac{l}{z}\,dv\, .
\ee
This choice is related to the canonical splitting \eqref{eq:KS-data} through the rescaling \eqref{eq:ambiguity} with $a(x) = z/l$, which leaves the metric \eqref{eq:black-string-EF} invariant. The vector $k_M$ is null and geodesic with respect to the AdS$_5$ background, and the determinant satisfies $\det(g) = \det(\bar{g})$, confirming that \eqref{eq:warped-KS-data} provides an equally valid Kerr--Schild representation.

Despite this equivalence at the gravitational level, the redistribution of the warp factor between $\phi$ and $k_M$ leads to a physically distinct realization of the classical double copy, as we now show.

\medskip
\noindent\textbf{Single copy:} The single copy gauge field now reads
\be
\label{eq:warped-A}
\tilde{A}_v = \frac{2Ml}{rz}\, ,
\ee
with all other components vanishing. The field strength has two independent non-vanishing components,
\be
\label{eq:warped-F}
\tilde{F}_{vr} = \frac{2Ml}{r^2 z}\, , \qquad \tilde{F}_{vz} = \frac{2Ml}{rz^2}\,,
\ee
in contrast with the canonical single copy, where the gauge field is $z$-independent and $F_{MN}$ is purely four-dimensional. Repeating the procedure of Section \ref{subsec:single-copy-CPT}, \textit{i.e.} contracting the linearized Einstein equation with $\xi = \partial_v$ (which now satisfies $\xi^{M}k_{M}=l/z$) and evaluating the Christoffel symbols of the AdS$_5$ background, one finds
\be
\label{eq:warped-Maxwell-source}
\bar{\nabla}_N \tilde{F}_M{}^N = \tilde{J}_M\,,
\ee
with source current
\be
\label{eq:warped-current}
\tilde{J}_M = -\frac{2M}{lr}\!\left(\frac{5}{z}\,\delta_M^v + \frac{1}{r}\,\delta_M^z\right).
\ee
Away from $r = 0$, this source is non-vanishing and extends along the holographic direction, in sharp contrast with the sourceless equation satisfied by the canonical single copy. The current is conserved, $\bar{\nabla}^M \tilde{J}_M = 0$, so the alternative splitting is internally consistent as a gauge theory, but the delocalized nature of the source signals that it does not admit a brane-localized interpretation.

The two single copies,
\be
A_v = \frac{2M}{r}\, , \qquad \tilde{A}_v = \frac{2Ml}{rz}\, ,
\ee
are not related by a gauge transformation. Indeed, their difference
\be
A_v - \tilde{A}_v = \frac{2M}{r}\left(1 - \frac{l}{z}\right)\, , 
\ee
would need to equal $\partial_v\Lambda$ for some scalar $\Lambda$, implying
\be
\Lambda = \frac{2M}{r}\left(1 - \frac{l}{z}\right)v\, , 
\ee
but then $\partial_r\Lambda \neq 0$, contradicting the fact that both fields have $A_r = 0$. Moreover, they satisfy different equations of motion: the canonical field is sourceless (away from $r = 0$), while the alternative one is supported by a delocalized bulk current.

\medskip
\noindent\textbf{Zeroth copy:}
The scalar associated with the alternative splitting is
\be
\tilde{\Phi} = \frac{2M}{r}\,,
\ee
which is independent of the holographic coordinate. Repeating the double contraction of Section \ref{subsec:zeroth-copy-CPT}, one finds
\be
\label{eq:alt-scalar-eom}
\left[\bar{\Box} - \frac{12}{l^2}\right]\left(\frac{l^2}{z^2}\tilde{\Phi}\right) = 0\, .
\ee
In contrast with the canonical zeroth copy equation \eqref{eq:scalar-warped}, eq. \eqref{eq:alt-scalar-eom} contains no first-derivative term along~$z$, reflecting the $z$-independence of~$\tilde{\Phi}$. The equation of motion for the scalar field reduces to the massless Klein--Gordon equation
\be
\label{eq:alt-scalar-eom-KG}
\bar{\Box}\tilde{\Phi} = 0\, .
\ee

The physical content of the two zeroth copies is markedly different. The canonical scalar $\Phi \sim z^2$ encodes the warp factor and, after the rescaling $\Theta \sim z^{-2}$, yields a normalizable mode localized near the brane satisfying a massive Klein--Gordon equation with $m^2 = 12/l^2$. The alternative scalar $\tilde{\Phi}$ is $z$-independent and therefore not normalizable with respect to the bulk norm\footnote{The norm $\int_l^\infty dz\,(l/z)^5\,|\tilde{\Phi}|^2 \sim\int_l^\infty dz\,z^{-5}$ converges due to the AdS$_5$ measure and not to a non-trivial decay of the field profile, which carries no information about the warped geometry. In contrast, the canonical $\Phi \sim z^2$ does encode the warp factor, and its rescaled version $\Theta \sim z^{-2}$ is normalizable.}: although it satisfies the massless equation $\bar{\Box}\,\tilde{\Phi} = 0$, the $z$-dependence has been entirely stripped from the Kerr--Schild scalar, so that the zeroth copy retains no imprint of the extra dimension. The massive Klein--Gordon equation satisfied by $(l^2/z^2)\tilde{\Phi}$ is not a property of the alternative splitting itself, but rather a consequence of the same rescaling that defines $\Theta$ in the canonical construction.

Taken together, these results confirm that the physicality criterion of \cite{Carrillo-Gonzalez:2017iyj}, requiring the absence of delocalized sources in the single copy and the retention of the warped bulk structure in the zeroth copy, selects the canonical splitting \eqref{eq:KS-data} as the physically preferred one. Different Kerr--Schild decompositions that are gravitationally equivalent encode different aspects of the warped geometry, and only the canonical choice leads to a double copy with a sourceless gauge field and a zeroth copy that carries a non-trivial imprint of the extra dimension.

\section{Conclusions}
\label{sec:conclusions}

In this work we have constructed the Kerr--Schild classical double copy of the RSII black string in five dimensions, using the Poincar\'e patch of AdS$_5$ as the background. The canonical splitting of~\cite{Carrillo-Gonzalez:2017iyj} yields a single copy gauge field that is independent of the holographic coordinate and satisfies a sourceless Maxwell equation on the curved background, in direct analogy with the Coulomb solution of the Schwarzschild double copy~\cite{Monteiro:2014cda}. The zeroth copy scalar obeys a modified Klein--Gordon equation with a first-order derivative coupling along the extra dimension; a field redefinition removes this coupling and produces a standard Klein--Gordon equation with effective mass $m^2 = 12/l^2$, with the redefined field $\Theta \sim z^{-2}$ normalizable and localized near the brane. An alternative Kerr--Schild splitting, gravitationally equivalent to the canonical one, leads to a physically inequivalent double copy: the gauge field is supported by a conserved but delocalized bulk current, and the zeroth copy carries no imprint of the warped geometry. These results demonstrate that the Kerr--Schild double copy is sensitive to the warped bulk structure of braneworld models, and that the physicality criterion of~\cite{Carrillo-Gonzalez:2017iyj}---requiring the absence of delocalized sources---plays an essential role in selecting the physically meaningful decomposition.

Our construction complements the recent analysis of~\cite{Alencar:2026zdz}, which studies the double copy of black strings in a four-dimensional cylindrical AdS background. While both works find gauge fields satisfying Maxwell-type equations on curved backgrounds, the five-dimensional RSII setting introduces qualitatively new features: the exponential localization of gravity along the extra dimension leaves a direct imprint on the gauge-theory description through the modified Klein--Gordon equation and the normalizability structure of the zeroth copy, which have no counterpart in that work.

Several directions remain open. A natural extension is to consider rotating and charged braneworld black holes~\cite{Aliev:2005bi,Neves:2012it}; in particular, the tidal charge inherited from the bulk Weyl tensor in~\cite{Aliev:2005bi} offers a concrete probe of how extra-dimensional effects are encoded in the single copy. It would also be interesting to investigate whether the double copy can track the Gregory--Laflamme instability~\cite{Gregory:1993vy} of the RSII black string by comparing the gauge-theory data of the unstable solution with that of its expected stable endpoint, and whether the construction extends to the stabilized black strings of~\cite{Fichet:2026ace}. From a holographic perspective, the effective mass $m^2 = 12/l^2$ corresponds via the standard AdS/CFT relation~\cite{Maldacena:1997re,Gubser:1998bc,Witten:1998qj} to a dual operator of conformal dimension $\Delta = 6$; understanding whether the single and zeroth copies admit a natural boundary CFT interpretation would connect the double copy programme to the holographic description of braneworld gravity~\cite{Karch:2000ct}. Finally, the Petrov type~D character of the black string makes it a natural candidate for the Weyl double copy~\cite{Luna:2018dpt}. In particular, recent progress has extended the Weyl double copy construction beyond four dimensions, including five-dimensional type~N and type~D spacetimes~\cite{Zhao:2024ljb,Zhao:2024wtn}. It would therefore be interesting to investigate whether the warped bulk geometry considered here admits a corresponding Weyl double copy description, or whether the Randall--Sundrum warping introduces genuinely new features. We leave these questions for future work.

\section*{Acknowledgements}

The author thanks Juan La Cruz for valuable discussions on braneworld physics that inspired this project. This work was supported by the UADE Academic Research Project \emph{``Black Hole Physics in Modern Theories of Gravity''} (A26T65).

\bibliography{references}

@article{Bern:2008qj,
    author = "Bern, Z. and Carrasco, J. J. M. and Johansson, Henrik",
    title = "{New Relations for Gauge-Theory Amplitudes}",
    eprint = "0805.3993",
    archivePrefix = "arXiv",
    primaryClass = "hep-ph",
    reportNumber = "UCLA-07-TEP-15",
    doi = "10.1103/PhysRevD.78.085011",
    journal = "Phys. Rev. D",
    volume = "78",
    pages = "085011",
    year = "2008"
}

@article{Bern:2010ue,
    author = "Bern, Zvi and Carrasco, John Joseph M. and Johansson, Henrik",
    title = "{Perturbative Quantum Gravity as a Double Copy of Gauge Theory}",
    eprint = "1004.0476",
    archivePrefix = "arXiv",
    primaryClass = "hep-th",
    reportNumber = "UCLA-10-TEP-102, SACLAY-IPHT-T10-044",
    doi = "10.1103/PhysRevLett.105.061602",
    journal = "Phys. Rev. Lett.",
    volume = "105",
    pages = "061602",
    year = "2010"
}

@article{Monteiro:2014cda,
    author = "Monteiro, Ricardo and O'Connell, Donal and White, Chris D.",
    title = "{Black holes and the double copy}",
    eprint = "1410.0239",
    archivePrefix = "arXiv",
    primaryClass = "hep-th",
    reportNumber = "EDINBURGH-2014-18",
    doi = "10.1007/JHEP12(2014)056",
    journal = "JHEP",
    volume = "12",
    number = "12",
    pages = "056",
    year = "2014"
}

@article{Kerr:1965wfc,
    author = "Kerr, Roy Patrick and Schild, Alfred",
    title = "{Some algebraically degenerate solutions of Einstein{\textquoteright}s gravitational field equations}",
    journal = "Proc. Symp. Appl. Math.",
    volume = "17",
    pages = "199",
    year = "1965"
}

@article{Luna:2015paa,
    author = "Luna, Andr{\'e}s and Monteiro, Ricardo and O'Connell, Donal and White, Chris D.",
    title = "{The classical double copy for Taub{\textendash}NUT spacetime}",
    eprint = "1507.01869",
    archivePrefix = "arXiv",
    primaryClass = "hep-th",
    doi = "10.1016/j.physletb.2015.09.021",
    journal = "Phys. Lett. B",
    volume = "750",
    pages = "272--277",
    year = "2015"
}

@article{Bahjat-Abbas:2017htu,
    author = "Bahjat-Abbas, Nadia and Luna, Andr{\'e}s and White, Chris D.",
    title = "{The Kerr-Schild double copy in curved spacetime}",
    eprint = "1710.01953",
    archivePrefix = "arXiv",
    primaryClass = "hep-th",
    reportNumber = "QMUL-PH-17-16",
    doi = "10.1007/JHEP12(2017)004",
    journal = "JHEP",
    volume = "12",
    number = "12",
    pages = "004",
    year = "2017"
}

@article{Carrillo-Gonzalez:2017iyj,
    author = "Carrillo-Gonz{\'a}lez, Mariana and Penco, Riccardo and Trodden, Mark",
    title = "{The classical double copy in maximally symmetric spacetimes}",
    eprint = "1711.01296",
    archivePrefix = "arXiv",
    primaryClass = "hep-th",
    doi = "10.1007/JHEP04(2018)028",
    journal = "JHEP",
    volume = "04",
    number = "04",
    pages = "028",
    year = "2018"
}

@article{Alkac:2021bav,
    author = "Alkac, Gokhan and Gumus, Mehmet Kemal and Tek, Mustafa",
    title = "{The Kerr-Schild Double Copy in Lifshitz Spacetime}",
    eprint = "2103.06986",
    archivePrefix = "arXiv",
    primaryClass = "hep-th",
    doi = "10.1007/JHEP05(2021)214",
    journal = "JHEP",
    volume = "05",
    number = "05",
    pages = "214",
    year = "2021"
}

@misc{Alencar:2026zdz,
    author = "Alencar, G. and Muniz, C. R. and Oliveira, M. S.",
    title = "{Classical double copy of black strings in an Anti-de Sitter background}",
    eprint = "2601.22383",
    archivePrefix = "arXiv",
    primaryClass = "hep-th",
    year = "2026"
}

@misc{Morieri:2026gdo,
    author = "Morieri, Riccardo and Pesando, Igor and Reichenberg Ashby, Michael L. and White, Chris D.",
    title = "{Classical strings and the double copy}",
    eprint = "2602.10907",
    archivePrefix = "arXiv",
    primaryClass = "hep-th",
    year = "2026"
}

@article{Lee:2018gxc,
    author = "Lee, Kanghoon",
    title = "{Kerr-Schild Double Field Theory and Classical Double Copy}",
    eprint = "1807.08443",
    archivePrefix = "arXiv",
    primaryClass = "hep-th",
    doi = "10.1007/JHEP10(2018)027",
    journal = "JHEP",
    volume = "10",
    number = "10",
    pages = "027",
    year = "2018"
}

@article{Cho:2019ype,
    author = "Cho, Wonyoung and Lee, Kanghoon",
    title = "{Heterotic Kerr-Schild Double Field Theory and Classical Double Copy}",
    eprint = "1904.11650",
    archivePrefix = "arXiv",
    primaryClass = "hep-th",
    doi = "10.1007/JHEP07(2019)030",
    journal = "JHEP",
    volume = "07",
    number = "07",
    pages = "030",
    year = "2019"
}

@article{Lescano:2020nve,
    author = "Lescano, Eric and Rodr\'\i{}guez, Jes\'us A.",
    title = "{$ \mathcal{N} $ = 1 supersymmetric Double Field Theory and the generalized Kerr-Schild ansatz}",
    eprint = "2002.07751",
    archivePrefix = "arXiv",
    primaryClass = "hep-th",
    doi = "10.1007/JHEP10(2020)148",
    journal = "JHEP",
    volume = "10",
    number = "10",
    pages = "148",
    year = "2020"
}

@article{Lescano:2021ooe,
    author = "Lescano, Eric and Rodr\'\i{}guez, Jes\'us A.",
    title = "{Higher-derivative heterotic Double Field Theory and classical double copy}",
    eprint = "2101.03376",
    archivePrefix = "arXiv",
    primaryClass = "hep-th",
    doi = "10.1007/JHEP07(2021)072",
    journal = "JHEP",
    volume = "07",
    number = "07",
    pages = "072",
    year = "2021"
}

@article{Angus:2021zhy,
    author = "Angus, Stephen and Cho, Kyoungho and Lee, Kanghoon",
    title = "{The classical double copy for half-maximal supergravities and T-duality}",
    eprint = "2105.12857",
    archivePrefix = "arXiv",
    primaryClass = "hep-th",
    reportNumber = "APCTP Pre2021 - 010",
    doi = "10.1007/JHEP10(2021)211",
    journal = "JHEP",
    volume = "10",
    number = "10",
    pages = "211",
    year = "2021"
}

@article{Lescano:2022nhp,
    author = "Lescano, Eric and Roychowdhury, Sourav",
    title = "{Heterotic Kerr-Schild Double Field Theory and its double Yang-Mills formulation}",
    eprint = "2201.09364",
    archivePrefix = "arXiv",
    primaryClass = "hep-th",
    doi = "10.1007/JHEP04(2022)090",
    journal = "JHEP",
    volume = "04",
    number = "04",
    pages = "090",
    year = "2022"
}

@article{Rodriguez:2025hiu,
    author = "Rodriguez, Jesus A.",
    title = "{Supersymmetric {\ensuremath{\alpha}}'-corrections to the generalized Kerr-Schild ansatz}",
    eprint = "2510.15130",
    archivePrefix = "arXiv",
    primaryClass = "hep-th",
    doi = "10.1007/JHEP01(2026)038",
    journal = "JHEP",
    volume = "01",
    number = "01",
    pages = "038",
    year = "2026"
}

@article{Berman:2020xvs,
    author = "Berman, David S. and Kim, Kwangeon and Lee, Kanghoon",
    title = "{The classical double copy for M-theory from a Kerr-Schild ansatz for exceptional field theory}",
    eprint = "2010.08255",
    archivePrefix = "arXiv",
    primaryClass = "hep-th",
    reportNumber = "QMUL-PH-20-27, APCTP Pre2020 - 025",
    doi = "10.1007/JHEP04(2021)071",
    journal = "JHEP",
    volume = "04",
    number = "04",
    pages = "071",
    year = "2021"
}

@misc{Berman:2022bgy,
    author = "Berman, David S. and Kim, Kwangeon and Lee, Kanghoon",
    title = "{Double copying Exceptional Field theories}",
    eprint = "2201.10854",
    archivePrefix = "arXiv",
    primaryClass = "hep-th",
    year = "2022"
}

@article{Randall:1999ee,
    author = "Randall, Lisa and Sundrum, Raman",
    title = "{A Large mass hierarchy from a small extra dimension}",
    eprint = "hep-ph/9905221",
    archivePrefix = "arXiv",
    reportNumber = "MIT-CTP-2860, PUPT-1860, BUHEP-99-9",
    doi = "10.1103/PhysRevLett.83.3370",
    journal = "Phys. Rev. Lett.",
    volume = "83",
    pages = "3370--3373",
    year = "1999"
}

@article{Randall:1999vf,
    author = "Randall, Lisa and Sundrum, Raman",
    title = "{An Alternative to compactification}",
    eprint = "hep-th/9906064",
    archivePrefix = "arXiv",
    reportNumber = "MIT-CTP-2874, PUPT-1867, BUHEP-99-13",
    doi = "10.1103/PhysRevLett.83.4690",
    journal = "Phys. Rev. Lett.",
    volume = "83",
    pages = "4690--4693",
    year = "1999"
}

@article{Garriga:1999yh,
    author = "Garriga, Jaume and Tanaka, Takahiro",
    title = "{Gravity in the brane world}",
    eprint = "hep-th/9911055",
    archivePrefix = "arXiv",
    reportNumber = "UAB-FT-476, OU-TAP-106",
    doi = "10.1103/PhysRevLett.84.2778",
    journal = "Phys. Rev. Lett.",
    volume = "84",
    pages = "2778--2781",
    year = "2000"
}

@article{Chamblin:1999by,
    author = "Chamblin, A. and Hawking, S. W. and Reall, H. S.",
    title = "{Brane world black holes}",
    eprint = "hep-th/9909205",
    archivePrefix = "arXiv",
    reportNumber = "DAMTP-1999-133",
    doi = "10.1103/PhysRevD.61.065007",
    journal = "Phys. Rev. D",
    volume = "61",
    pages = "065007",
    year = "2000"
}

@article{Dadhich:2000am,
    author = "Dadhich, Naresh and Maartens, Roy and Papadopoulos, Philippos and Rezania, Vahid",
    title = "{Black holes on the brane}",
    eprint = "hep-th/0003061",
    archivePrefix = "arXiv",
    doi = "10.1016/S0370-2693(00)00798-X",
    journal = "Phys. Lett. B",
    volume = "487",
    pages = "1--6",
    year = "2000"
}

@article{Tanaka:2002rb,
    author = "Tanaka, Takahiro",
    editor = "Maeda, K. and Sasaki, M.",
    title = "{Classical black hole evaporation in Randall-Sundrum infinite brane world}",
    eprint = "gr-qc/0203082",
    archivePrefix = "arXiv",
    doi = "10.1143/PTPS.148.307",
    journal = "Prog. Theor. Phys. Suppl.",
    volume = "148",
    pages = "307--316",
    year = "2003"
}

@article{Emparan:2002px,
    author = "Emparan, Roberto and Fabbri, Alessandro and Kaloper, Nemanja",
    title = "{Quantum black holes as holograms in AdS brane worlds}",
    eprint = "hep-th/0206155",
    archivePrefix = "arXiv",
    reportNumber = "SU-ITP-02-23, CERN-TH-2002-131",
    doi = "10.1088/1126-6708/2002/08/043",
    journal = "JHEP",
    volume = "08",
    number = "08",
    pages = "043",
    year = "2002"
}

@article{Aliev:2005bi,
    author = "Aliev, A. N. and Gumrukcuoglu, A. E.",
    title = "{Charged rotating black holes on a 3-brane}",
    eprint = "hep-th/0502223",
    archivePrefix = "arXiv",
    doi = "10.1103/PhysRevD.71.104027",
    journal = "Phys. Rev. D",
    volume = "71",
    pages = "104027",
    year = "2005"
}

@article{Amarilla:2011fx,
    author = "Amarilla, Leonardo and Eiroa, Ernesto F.",
    title = "{Shadow of a rotating braneworld black hole}",
    eprint = "1112.6349",
    archivePrefix = "arXiv",
    primaryClass = "gr-qc",
    doi = "10.1103/PhysRevD.85.064019",
    journal = "Phys. Rev. D",
    volume = "85",
    pages = "064019",
    year = "2012"
}

@article{Neves:2012it,
    author = "Neves, J. C. S. and Molina, C.",
    title = "{Rotating black holes in a Randall-Sundrum brane with a cosmological constant}",
    eprint = "1211.2848",
    archivePrefix = "arXiv",
    primaryClass = "gr-qc",
    doi = "10.1103/PhysRevD.86.124047",
    journal = "Phys. Rev. D",
    volume = "86",
    pages = "124047",
    year = "2012"
}

@article{Gregory:1993vy,
    author = "Gregory, R. and Laflamme, R.",
    title = "{Black strings and p-branes are unstable}",
    eprint = "hep-th/9301052",
    archivePrefix = "arXiv",
    doi = "10.1103/PhysRevLett.70.2837",
    journal = "Phys. Rev. Lett.",
    volume = "70",
    pages = "2837--2840",
    year = "1993"
}

@article{Chamblin:1999ya,
    author = "Chamblin, H. A. and Reall, H. S.",
    title = "{Dynamic dilatonic domain walls}",
    eprint = "hep-th/9903225",
    archivePrefix = "arXiv",
    reportNumber = "DAMTP-1999-35",
    doi = "10.1016/S0550-3213(99)00520-9",
    journal = "Nucl. Phys. B",
    volume = "562",
    pages = "133--157",
    year = "1999"
}

@article{Brecher:1999xf,
    author = "Brecher, D. and Perry, M. J.",
    title = "{Ricci flat branes}",
    eprint = "hep-th/9908018",
    archivePrefix = "arXiv",
    reportNumber = "DAMTP-1999-97",
    doi = "10.1016/S0550-3213(99)00659-8",
    journal = "Nucl. Phys. B",
    volume = "566",
    pages = "151--172",
    year = "2000"
}

@article{Israel:1966rt,
    author = "Israel, W.",
    title = "{Singular hypersurfaces and thin shells in general relativity}",
    doi = "10.1007/BF02710419",
    journal = "Nuovo Cim. B",
    volume = "44S10",
    pages = "1",
    year = "1966",
    note = "[Erratum: Nuovo Cim.B 48, 463 (1967)]"
}

@article{Luna:2018dpt,
    author = "Luna, Andr{\'e}s and Monteiro, Ricardo and Nicholson, Isobel and O'Connell, Donal",
    title = "{Type D Spacetimes and the Weyl Double Copy}",
    eprint = "1810.08183",
    archivePrefix = "arXiv",
    primaryClass = "hep-th",
    doi = "10.1088/1361-6382/ab03e6",
    journal = "Class. Quant. Grav.",
    volume = "36",
    pages = "065003",
    year = "2019"
}

@article{Davoudiasl:1999tf,
    author = "Davoudiasl, H. and Hewett, J. L. and Rizzo, T. G.",
    title = "{Bulk gauge fields in the Randall-Sundrum model}",
    eprint = "hep-ph/9911262",
    archivePrefix = "arXiv",
    reportNumber = "SLAC-PUB-8298",
    doi = "10.1016/S0370-2693(99)01430-6",
    journal = "Phys. Lett. B",
    volume = "473",
    pages = "43--49",
    year = "2000"
}

@article{Pomarol:1999ad,
    author = "Pomarol, Alex",
    title = "{Gauge bosons in a five-dimensional theory with localized gravity}",
    eprint = "hep-ph/9911294",
    archivePrefix = "arXiv",
    reportNumber = "CERN-TH-99-341",
    doi = "10.1016/S0370-2693(00)00737-1",
    journal = "Phys. Lett. B",
    volume = "486",
    pages = "153--157",
    year = "2000"
}

@article{Ghoroku:2001zu,
    author = "Ghoroku, Kazuo and Nakamura, Akihiro",
    title = "{Massive vector trapping as a gauge boson on a brane}",
    eprint = "hep-th/0106145",
    archivePrefix = "arXiv",
    reportNumber = "FIT-HE-01-02, KAGOSHIMA-HE-01-3",
    doi = "10.1103/PhysRevD.65.084017",
    journal = "Phys. Rev. D",
    volume = "65",
    pages = "084017",
    year = "2002"
}

@misc{Alencar:2026afq,
    author = "Alencar, G. and Almeida, R. S. and Costa Filho, R. N. and Crispim, T. M. and Lobo, Francisco S. N.",
    title = "{The End of the Road for Bulk Fields in Braneworlds}",
    eprint = "2601.05190",
    archivePrefix = "arXiv",
    primaryClass = "gr-qc",
    year = "2026"
}

@misc{Fichet:2026ace,
    author = "Fichet, Sylvain and Megias, Eugenio and Quiros, Mariano and Yamanaki, Geovanna",
    title = "{Stable Black Strings from Warped Backgrounds}",
    eprint = "2603.12332",
    archivePrefix = "arXiv",
    primaryClass = "hep-th",
    year = "2026"
}

@article{Maldacena:1997re,
    author = "Maldacena, Juan Martin",
    title = "{The Large $N$ limit of superconformal field theories and supergravity}",
    eprint = "hep-th/9711200",
    archivePrefix = "arXiv",
    reportNumber = "HUTP-97-A097, HUTP-98-A097",
    doi = "10.4310/ATMP.1998.v2.n2.a1",
    journal = "Adv. Theor. Math. Phys.",
    volume = "2",
    pages = "231--252",
    year = "1998"
}

@article{Gubser:1998bc,
    author = "Gubser, S. S. and Klebanov, Igor R. and Polyakov, Alexander M.",
    title = "{Gauge theory correlators from noncritical string theory}",
    eprint = "hep-th/9802109",
    archivePrefix = "arXiv",
    reportNumber = "PUPT-1767",
    doi = "10.1016/S0370-2693(98)00377-3",
    journal = "Phys. Lett. B",
    volume = "428",
    pages = "105--114",
    year = "1998"
}

@article{Witten:1998qj,
    author = "Witten, Edward",
    title = "{Anti de Sitter space and holography}",
    eprint = "hep-th/9802150",
    archivePrefix = "arXiv",
    reportNumber = "IASSNS-HEP-98-15",
    doi = "10.4310/ATMP.1998.v2.n2.a2",
    journal = "Adv. Theor. Math. Phys.",
    volume = "2",
    pages = "253--291",
    year = "1998"
}

@article{Karch:2000ct,
    author = "Karch, Andreas and Randall, Lisa",
    editor = "Duff, Michael J. and Liu, J. T. and Lu, J.",
    title = "{Locally localized gravity}",
    eprint = "hep-th/0011156",
    archivePrefix = "arXiv",
    reportNumber = "MIT-CTP-3099",
    doi = "10.1088/1126-6708/2001/05/008",
    journal = "JHEP",
    volume = "05",
    number = "05",
    pages = "008",
    year = "2001"
}

@article{Zhao:2024ljb,
    author = "Zhao, Weicheng and Mao, Pu-Jian and Wu, Jun-Bao",
    title = "{Five dimensional Weyl double copy}",
    eprint = "2409.06786",
    archivePrefix = "arXiv",
    primaryClass = "hep-th",
    reportNumber = "USTC-ICTS/PCFT-24-30",
    doi = "10.1103/PhysRevD.111.L081902",
    journal = "Phys. Rev. D",
    volume = "111",
    number = "8",
    pages = "L081902",
    year = "2025"
}

@article{Zhao:2024wtn,
    author = "Zhao, Weicheng and Mao, Pu-Jian and Wu, Jun-Bao",
    title = "{Weyl double copy in type D spacetime in four and five dimensions}",
    eprint = "2411.04774",
    archivePrefix = "arXiv",
    primaryClass = "hep-th",
    reportNumber = "USTC-ICTS/PCFT-24-42",
    doi = "10.1103/PhysRevD.111.066005",
    journal = "Phys. Rev. D",
    volume = "111",
    number = "6",
    pages = "066005",
    year = "2025"
}

\end{document}